\documentclass[nofootinbib,tightenlines, floatfix,eqsecnum,superscriptaddress,showkeys]{svproc}

\usepackage[utf8]{inputenc}          % file encoding
\usepackage{graphicx,xcolor}         % Graphics
\usepackage{array,dcolumn,longtable} % Tables
\usepackage{amsmath,amssymb,slashed} % Math symbols
\usepackage{multicol,footmisc}         % Graphics

\providecommand\lfstyle{}                   % lining figures
\providecommand\romanup[1]{\text{#1}}       % upright roman
\renewcommand\textsc{\MakeUppercase}

\usepackage{siunitx}                                  % Parsing numbers and units
\DeclareSIUnit{\fm}{\femto\metre}                     % fm: femtometre

\newcommand\lhc{\textsc{lhc}}

\newcommand\cms{\textsc{cms}}

\newcommand\dabmod{\textsc{dab-mod}}
\newcommand\vusphydro{\text{v-\textsc{usp}hydro}}

\newcommand\fonll{\textsc{fonll}}

\declareslashed{}{\divslash}{0.08}{0}{\mathcal{D}}

\DeclareSIUnit{\fm}{\femto\metre}

\newcommand\proton{{\romanup{p}}}

\newcommand\Dmeson{{\romanup{D}}}

\newcommand\Dzero{{\Dmeson^0}}

\newcommand\pp{{\proton\proton}}

\newcommand\pPb{{\proton\romanup{Pb}}}

\newcommand\PbPb{{\romanup{PbPb}}}

\newcommand\ArAr{{\romanup{ArAr}}}
\newcommand\XeXe{{\romanup{XeXe}}}
\newcommand\OO{{\romanup{OO}}}

\newcommand\Xe{{\romanup{Xe}}}

\newcommand\snn[1][]{\sqrt{s_\text{NN}}\ifx\\#1\\\else=\SI{#1}{\TeV}\fi}

\newcommand\pt{p_\text{T}}

\newcommand\raa{R_\text{AA}}
\newcommand\vn[1]{v_{#1}}
\newcommand\vnn{\vn{n}}

\begin{document}
\mainmatter
\title{\mbox{$\Dmeson$ meson sensitivity to a system size scan at LHC.}}
\date{\today}

\author{Roland Katz\inst{1} \and  Jacquelyn Noronha-Hostler\inst{2} \and \\  Caio A.~G.~Prado\inst{3} \and Alexandre A.~P.~Suaide\inst{4}}

\authorrunning{Roland Katz et al.}

\institute{SUBATECH, Univ. de Nantes, IMT, IN2P3/CNRS, 44307 Nantes, France
\and Dept. of Physics, Univ. of Illinois at Urbana-Champaign, Urbana, IL 61801, USA
\and Inst. of Particle Phys., Central China Normal Univ., Wuhan, Hubei 430079, China
\and Inst. de F\'{i}sica, Univ. de S\~{a}o Paulo, C.P. 66318, 05315-970 S\~{a}o Paulo, SP, Brazil}

\maketitle

\begin{abstract}
\vspace{-4mm}
Experimental measurements in pA collisions indicate no $\Dmeson$ meson suppression ($R_\pPb \sim 1$) but a surprisingly large $v_2$.  To better understand these results we propose a system size scan at the \lhc\ involving $^{16}\OO$, $^{40}\ArAr$, $^{129}\XeXe$ and $^{208}\PbPb$ collisions. Using Trento+ \vusphydro+\dabmod\ to make predictions, we find that the $\raa$ tends towards unity when the system size is decreased, but nonetheless, in the most central collisions $\vn2\{2\}$ is almost independent of the colliding system. These results are analyzed in light of path length and initial eccentricity variations.
\end{abstract}
%\pacs{Find Pacs numbers}
%\keywords{Put keywords}

\noindent \textsl{1. Introduction.}  Flow correlations, strangeness enhancement and suppression of hard probes are considered to be three signatures of the Quark-Gluon Plasma. The recent observation of the first two in small hadronic collisions - such as $\pp$ and $\pPb$ - raises many questions on the nature of the created ``medium'' in these collisions \cite{Khachatryan:2016txc,ALICE:2017jyt,PHENIX:2018lia}. Jet and heavy flavor suppression is not observed in small systems \cite{Acharya:2019mno}, but the \cms\ collaboration has measured large $\Dmeson$ meson anisotropies in $\pPb$ collisions~\cite{Sirunyan:2018toe}. We still dot not understand of how such a significant $\vn2$ in small systems could be compatible with $\raa \rightarrow 1$ \cite{Xu:2015iha}. In order to determine the applicability of hydrodynamics in these tiny systems, it was recently proposed to run a system size scan at \lhc\ via $\ArAr$ and $\OO$ collisions~\cite{Citron:2018lsq}, on which various predictions have been made~\cite{Sievert:2019zjr}. $\Dmeson$ mesons being mostly sensitive to equilibrium dynamics, they appear to be ideal probes of system size effects \cite{Katz:2019qwv}. Here we investigate these effects on the $\raa$ and $\vnn\{2\}$ by varying the colliding nuclei. To do so, we use Trento+\vusphydro+\dabmod \,\cite{Prado:2016szr} with the same soft backgrounds as in~\cite{Sievert:2019zjr} and the Langevin set up that gave us the best results in $\PbPb$ collisions \cite{Katz:2019fkc}. 

\vspace{3mm}
\noindent \textsl{2. Model Description.} The Monte Carlo simulation \dabmod~\cite{Katz:2019qwv,Prado:2016szr,Katz:2019fkc}, developed to study open heavy favours, is coupled to 2D+1 event-by-event hydrodynamical backgrounds~\cite{Sievert:2019zjr}. Heavy quarks are first sampled using \fonll\ distributions and then propagate with a relativistic Langevin model using the spatial diffusion coefficient from~\cite{Moore:2004tg}. When the heavy quark / medium decoupling temperature $T_d$ is reached, the hadronization is finally performed using a hybrid fragmentation/coalescence model. In \dabmod \ the overall magnitude of the $\raa$ is ambiguous as we usually fix the scaling parameter of the Langevin model using $\raa$ data in most central collisions. Here, we use for all colliding systems the value determined in $\PbPb$ collisions (i.e. $D/(2\pi T)=2.2$). 

\vspace{-5mm}
\begin{figure}[ht!]
    \centering
    \includegraphics[width=1.01\textwidth]{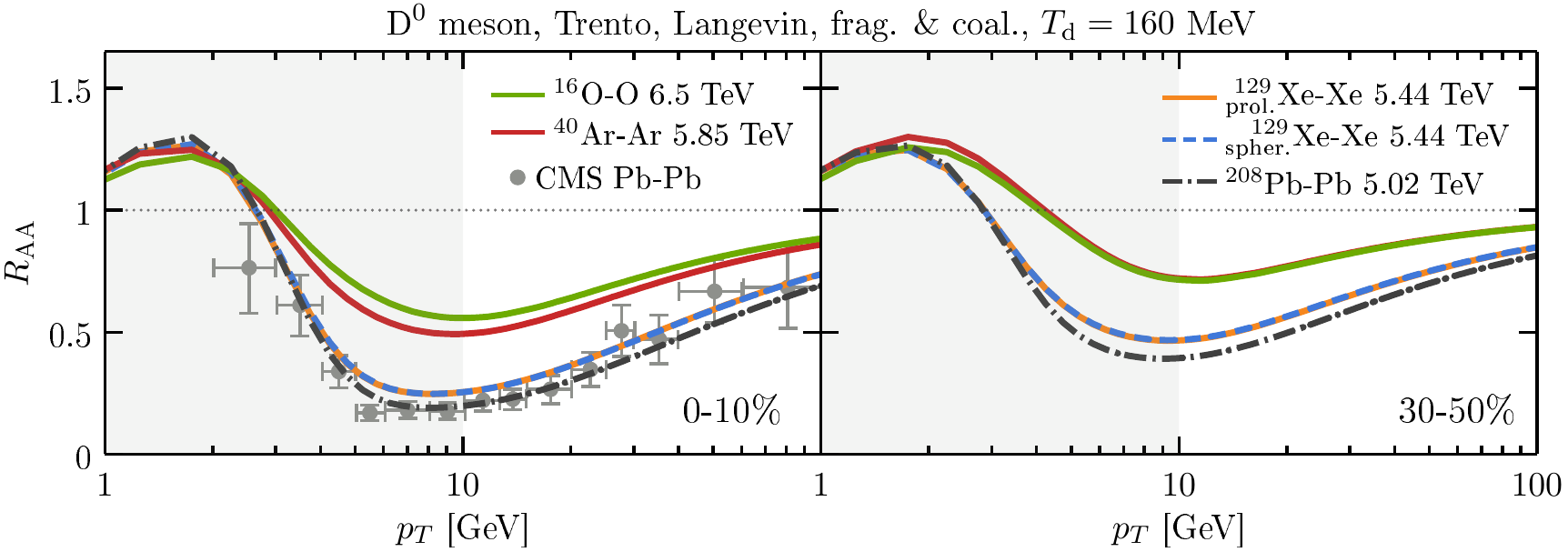}
    \caption{Direct $\Dzero$ meson $\raa$ for $\OO$, $\ArAr$, $\XeXe$ with spherical and prolate initial nuclei, and $\PbPb$ collisions in 0--10\% (left) and  30--50\% (right) centrality classes.  Prompt $\Dzero$ data ($|y|<1$) from the CMS collaboration for $\PbPb$ collisions~\cite{Sirunyan:2017xss}.}
    \label{fig:RAA}
\end{figure}
\vspace{-4.9mm}

\noindent \textsl{3. Results.} In Fig.~\ref{fig:RAA}, we first investigate how the system size modifies the $\raa$ as one moves towards smaller systems. First, central collisions are observed to be more sensitive to system size than mid-central collisions where there is no visible difference between $\OO$ and $\ArAr$ even though their system size is different~\cite{Sievert:2019zjr}. We expect the $\raa$ to smoothly approach unity with shrinking system size, as $(1-\raa)$ is approximatively proportional to the initial medium radius $\sim A^{1/3}$. However, in a future extension to $\pPb$ collisions, we might not reach unity enough to describe the data (like in previous studies with similar frameworks \cite{Xu:2015iha}). Finally, the deformation of the Xe nuclei has no influence on the $\raa$.

\begin{figure}[h!]
    \centering
    \includegraphics[width=0.49\textwidth]{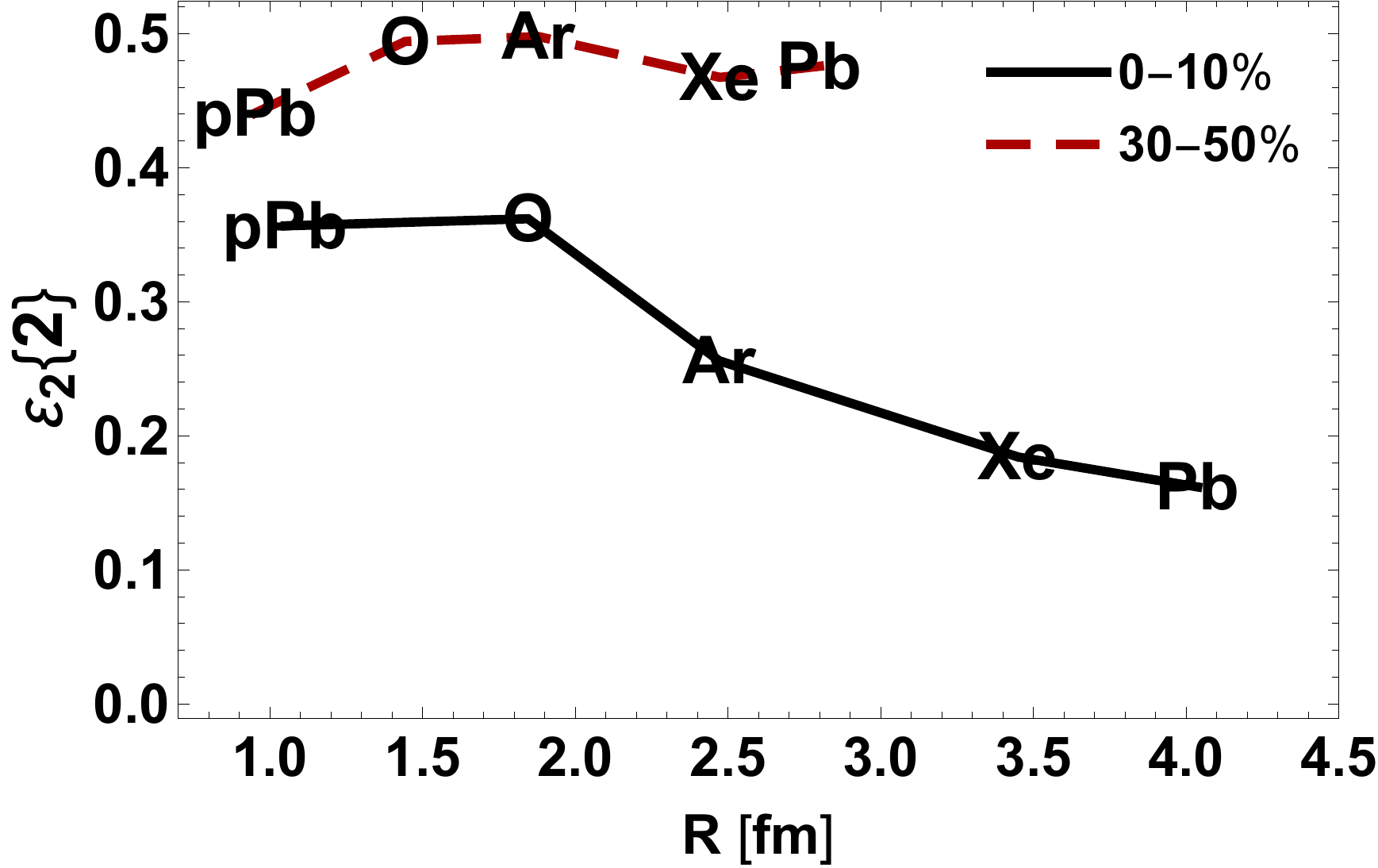}
%\hspace{1mm}
    \includegraphics[width=0.49\textwidth]{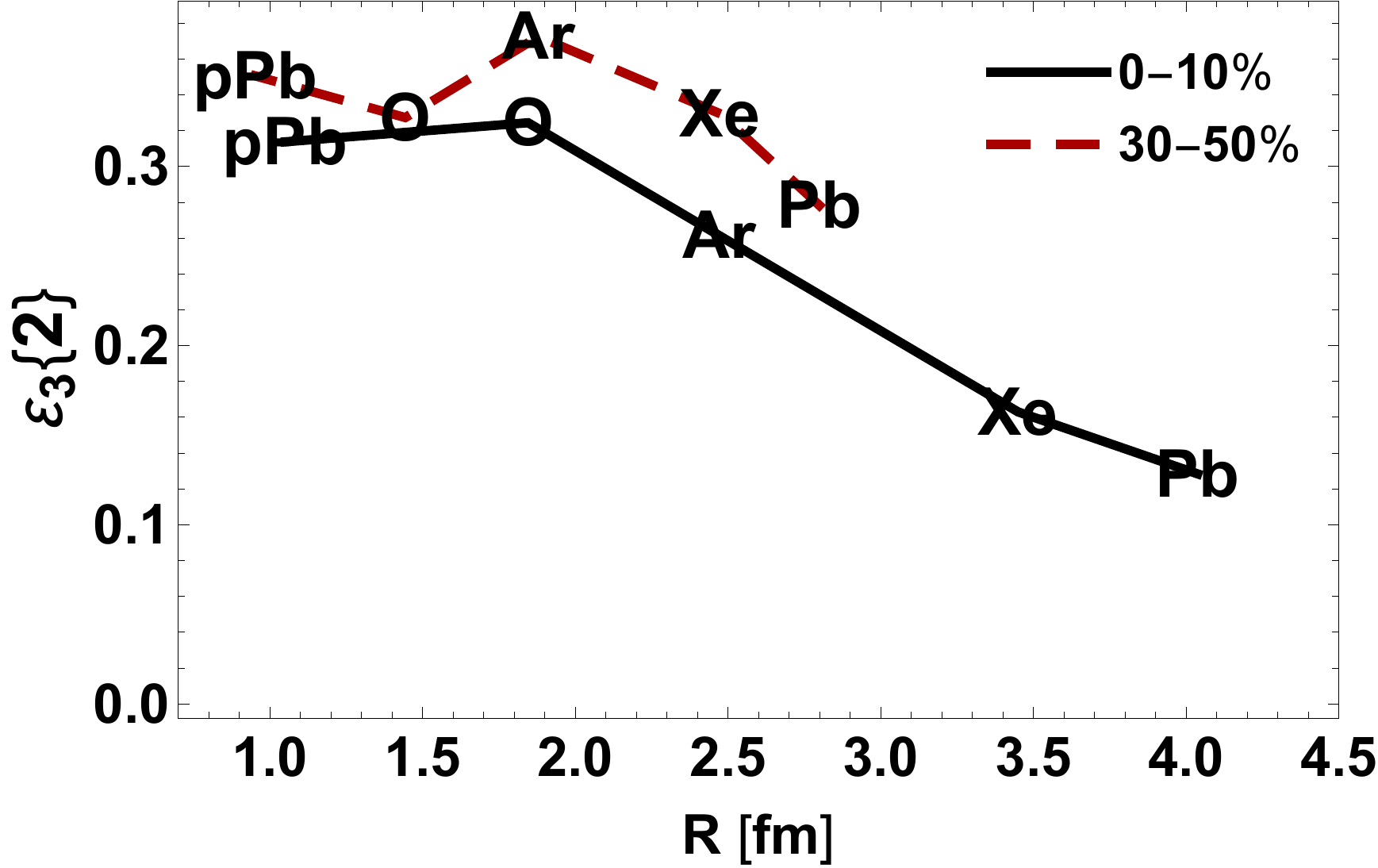}
    \caption{$\varepsilon_2\{2\}$ (left) and $\varepsilon_3\{2\}$ (right) versus radius for $\OO$, $\ArAr$, $\XeXe$ and $\PbPb$ collisions at the \lhc\ in 0--10\% and  30--50\% centrality classes.}
    \label{fig:ecc}    
\end{figure}

\begin{figure}[h!]
    \centering
    \includegraphics[width=1.01\textwidth]{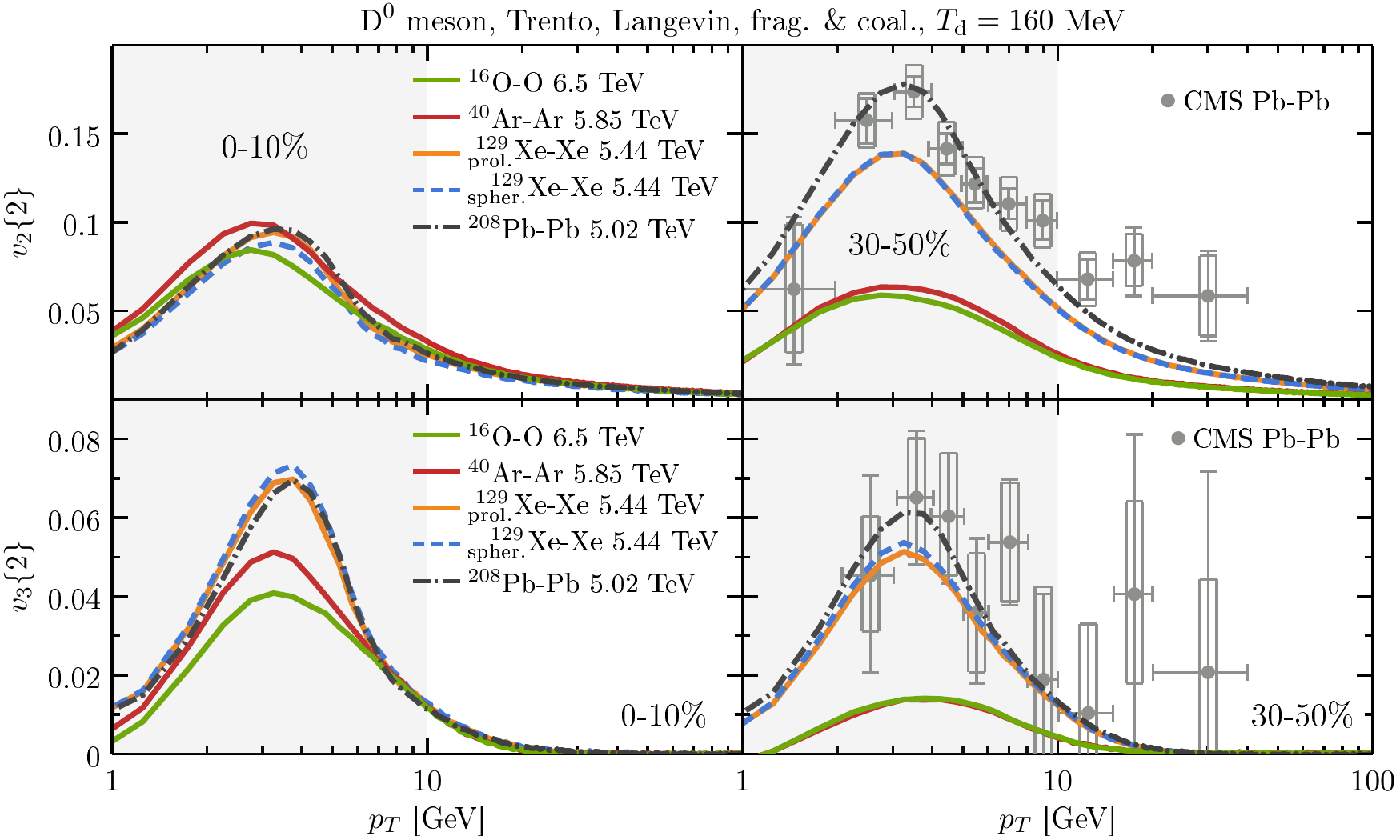}
    \caption{ Direct $\Dzero$ meson $\vn2\{2\}$ (top) and $\vn3\{2\}$ (bottom) for $\OO$, $\ArAr$, $\XeXe$ with spherical and prolate initial nuclei, and $\PbPb$ collisions in 0--10\% (left) and  30--50\% (right) centrality classes. Prompt $\Dzero$ data ($|y|<1$) from the CMS collaboration for $\PbPb$ collisions~\cite{Sirunyan:2017plt}.}
    \label{fig:v2}
\end{figure}

%\vspace{1mm}
\indent For the azimuthal anisotropies $\vnn$, two factors play a significant role: the size of the system, which can be described by the typical radius of the initial conditions R, and the initial geometrical shape usually characterized by the eccentricities $\varepsilon_n$~\cite{Sievert:2019zjr}. Their variations with the colliding system in the two centrality classes considered here are shown in Fig.~\ref{fig:ecc}. The systems resulting from $\OO$, $\ArAr$, $\XeXe$ and $\PbPb$ central collisions have significantly different sizes and eccentricities: the eccentricities increase while the radius decreases.  In contrast, in mid-central collisions the eccentricities remain roughly constant when one varies the system size. The mid-central collisions can then be seen as probes of pure system size effects. As measured $\Dmeson$ meson data in $\pPb$~\cite{Sirunyan:2018toe} correspond to central collisions, they might experience both system size and eccentricities variations compared to large AA collisions. In Fig.~\ref{fig:v2} (top) we  show the $\Dmeson$ meson $\vn2$ in the two different centrality classes. In the 30--50\% mid-central class, the $\vn2$ of smaller systems are significantly suppressed across all $\pt$. Thus, as $\varepsilon_2$ is $\sim \text{const.}$ in the mid-central class, the pure effect of the system size plays a dramatic role on the $\vn2$. In the 0--10\% centrality class, the $\vn2$ is observed to be roughly independent of the colliding system across all $\pt$. This striking result can be understood by returning to Fig.~\ref{fig:ecc} where we saw that for central collisions the $\varepsilon_2$ increases as R

\vspace{5mm}
\vspace{-12mm}

\vspace{-5mm}

\noindent  decreases. Thus, there are now two competing factors that influence the final $\vn2$: a suppression effect from decreasing R, like in the mid-central class, and an enhancement effect from increasing $\varepsilon_2$. The similarity of the $\vn2$ curves regardless of colliding system can therefore be explained by the two competing effects roughly compensating each other in central collisions. One can extend these ideas to $\pPb$ collisions: if they have large enough eccentricities (see Fig.~\ref{fig:ecc}) $\vn2$ may not vanish despite the system size shrinking (other effects, e.g. the initial flow, could also contribute). Note finally that in central collisions the $\vn2$ shows a sensitivity to the deformation of the $^{129}\Xe$ nucleus. In Fig.~\ref{fig:v2} (bottom), the triangular anisotropies $\vn3$ are observed to be more sensitive to size effects than by eccentricities, i.e. there is a consistent suppression in small systems regardless of the centrality class, even when $\varepsilon_3$ changes significantly. Finally, contrasting with the ``universality'' of $\vn3(\pt)$ across centralities usually observed in $\PbPb$ collisions~\cite{Sirunyan:2017plt}, in smaller systems the $\vn3$ vary strongly with the centrality class.

\vspace{3mm}
\noindent \textsl{4. Conclusions.} We made predictions for the $\Dmeson$ meson nuclear modification factors and azimuthal anisotropies for the proposed system size scan at \lhc~\cite{Katz:2019qwv}. We find that the $\raa$ gradually approaches unity as the system size decreases, i.e. as the path lengths shrink. The variations of the $\vn2$ over the colliding systems depend strongly on two competing factors: the typical system radius R and the geometry of the initial condition described by its eccentricity $\varepsilon_2$. In mid-central collisions, we get a clear hierarchy of the $\vn2$ between colliding systems, showing the strong influence of the system size itself, as $\varepsilon_2$ is nearly constant over the different systems. In central collisions the suppression of $\vn2$ due to the decreasing R is counterbalanced by an enhancement coming from an increasing $\varepsilon_2$, leading to roughly equivalent $\vn2(\pt)$ across the colliding system scan. Although $\varepsilon_3$ increases with decreasing R in central collisions, $\vn3$ is more sensitive to R itself and, thus, one observes a suppression following the system size hierarchy regardless of the centrality class. Finally, we find that in small systems $\vn3$ decreases with centrality, whereas it is known to be almost constant in $\PbPb$ collisions. The latter can now be explained by a balance between a suppression effect from path length reduction and an enhancement from $\varepsilon_3$ increase with centrality. 

\vspace{3mm}
\noindent \textsl{Acknowledgments.} The authors thank Funda\c{c}\~ao de Amparo \`a Pesquisa do Estado de S\~ao Paulo (FAPESP) and Conselho Nacional de Desenvolvimento Cient\'ifico e Tecnol\'ogico (CNPq) for support. R.K. is supported by the Region Pays de la Loire (France) under contract No. 2015-08473. C.A.G.P. is supported by the NSFC under grant No. 11890714 and 1186113100, MOST of China under Project No. 2014CB845404.  J.N.H. acknowledges the support of the Alfred P. Sloan Foundation and from the US-DOE Nuclear Science Grant No. DE-SC0019175.

\end{document}